\documentclass[12pt]{iopart}
\usepackage{graphicx} 
\usepackage{amssymb}
\usepackage{lipsum}
\usepackage{csquotes}

\expandafter\let\csname equation*\endcsname\relax
\expandafter\let\csname endequation*\endcsname\relax
\usepackage{amsmath}
\usepackage[dvipsnames]{xcolor}
\usepackage[noadjust]{cite}

\DeclareUnicodeCharacter{2212}{-}
\begin{document}
\sloppy

\title{Magnetization Dynamics in Quasiperiodic Magnonic Crystals}
\author{Riya Mehta$^1$, Bivas Rana$^2$$^\ast$, Susmita Saha$^1$$^\$$ }

\address{$^1$Department of Physics, Ashoka University, Sonipat, Haryana 131029, India

$^2$Institute of Spintronics and Quantum Information, Faculty of Physics, Adam Mickiewicz University in Poznan, Uniwersytetu Poznanskiego 2, 61-614 Poznan, Poland}
\ead{$^\ast$bivran@amu.edu.pl, $^\$$susmita.saha@ashoka.edu.in}
\vspace{10pt}
\begin{indented}
\item[]
\end{indented}

\begin{abstract}
Quasiperiodic magnonic crystals, in contrast to their periodic counterparts, lack strict periodicity which gives rise to complex and localised spin wave spectra characterized by numerous band gaps and fractal features. Despite their intrinsic structural complexity, quasiperiodic nature of these magnonic crystals enables better tunability of spin wave spectra over their periodic counterparts and therefore holds promise for the applications in reprogrammable magnonic devices. In this article, we provide an overview of magnetization reversal and precessional magnetization dynamics studied so far in various quasiperiodic magnonic crystals, illustrating how their quasiperiodic nature gives rise to tailored band structure, enabling unparalleled control over spin waves.  The review is concluded by highlighting the possible potential applications of these quasiperiodic magnonic crystals, exploring potential avenues for future exploration followed by a brief summary.
\end{abstract}

\section{Introduction}
The captivating world of unconventional quasiperiodic crystals unfolded in 1984 when Shechtman et al. made a groundbreaking discovery in a rapidly cooled aluminium manganese (Al-Mn) alloy \cite{shechtman1984metallic}. The observation of a quasiperiodic crystal with five-fold symmetry defied the long-held belief that crystals must exhibit translational periodicity. Quasiperiodic crystals possess long-range order without the constraints of translational periodicity, resulting in unique rotational symmetries and sharp diffraction patterns \cite{levine1984quasicrystals}. Since their initial discovery, quasiperiodic crystals have been found in a wide range of intermetallic alloys, further expanding the understanding and exploration of these intriguing materials \cite{steurer2004twenty}. 

 Due to the presence of unconventional rotational symmetries including five-fold, eight-fold and ten-fold, quasiperiodic crystals challenges traditional crystallographic theories \cite{janot1996quasicrystals}. Research in quasiperiodic photonic and phononic crystals has sparked significant interest due to their unique properties and potential applications. In the field of quasiperiodic photonic crystals, where electromagnetic waves serve as the means of communication, investigations have been focused on bandgap engineering \cite{wang2005photonic}. There the band structures are tailored to achieve enhanced photonic bandgaps for better control of light propagation and confinement. Moreover, these quasiperiodic structures have been studied for their ability to exhibit omnidirectional bandgaps, ensuring effective light confinement and filtering from all directions \cite{araujo2012omnidirectional}. Quasiperiodic photonic crystals have also shown promise in enhancing light-matter interaction, leading to potential applications in sensors \cite{shi2019refractive}, filters \cite{liu2018broadband,yan2018polarization}, lenses \cite{feng2005negative,zhang2018subwavelength}, prisms \cite{zhang2018non}, and optical fibres \cite{yan2018polarization,liu2018broadband}. Similarly, the exploration of quasiperiodic phononic crystals, where acoustic waves serve as the means of communication, has been conducted to advance the engineering of phononic bandgaps \cite{mehaney2019evolution}, enabling their applications in novel phononic devices \cite{kohmoto1987localization} and waveguides \cite{chen2012wave}. Additionally, the emergence of narrow band gaps in quasiperiodic phononic crystals as compared to periodic crystals offers their potential application in low frequency filters \cite{chen2008elastic}. These studies collectively demonstrate the versatility and potential of quasiperiodic photonic and phononic crystals in wave control, wave propagation, signal processing, sensing and applications in advanced materials  \cite{steurer2007photonic,imanian2021highly}.

Analogous to photonic and phononic crystals, magnonic crystals (MCs) could also be formed, where spin waves (SWs) serve as the communication media \cite{lenk2011building, kruglyak2010magnonics, neusser2009magnonics,kim2010micromagnetic,chumak2008scattering,vysotskii2005magnetostatic,chumak2022advances}. SWs are basically collective precessional motion of ordered magnetic spins coupled by short range exchange and long range dipolar interactions. When a magnetic media experiences an external perturbation, it minimizes exchange energy by realigning magnetization direction from their equilibrium orientations over a length scale, typically larger than the exchange length. This results in the propagation of magnetic disturbance in the form of SWs, with the fundamental quanta of SWs being magnons, which serve as carriers and processors of information. There are several ways to fabricate the MCs as discussed in reference \cite{chumak2017magnonic}. MCs can be developed by organizing magnetostatically coupled nanomagnets (e.g. magnetic dots, nanowires or nanostripes) in one, two or three-dimensional arrays or by making periodic arrays of holes in a continuous magnetic film. Sometimes magnetic islands with one material are embedded into a magnetic film made of another material to create MCs, known as bicomponent MCs \cite{choudhury2019anisotropic,ma2011micromagnetic}. The bicomponent MCs are also prepared by alternately arranging magnetic nanoelements made of two different magnetic materials \cite{gubbiotti2012collective}. In contrast to photonic and phononic crystals, the periodic potential in MCs can be modified by bias magnetic field strength and orientation, saturation magnetization and anisotropy of the magnetic media \cite{saha2013tunable}. Although magnonics shares similarities with photonics and phononics, yet it boasts distinct advantages compared to its photonic and phononic counterparts. These advantages encompass integrability with complementary metal-oxide semiconductor (CMOS) structures, programmability, smaller device footprints, anisotropic dispersion characteristics, presence of negative group velocity and shorter wavelengths down to few nanometer, which make MCs a suitable candidate for nanoscale on-chip communication devices, including magnonic waveguides \cite{demidov2011excitation,klos2012effect}, filters \cite{kim2009gigahertz}, splitters, phase shifters \cite{au2012nanoscale}, SWs emitters \cite{kaka2005mutual,yu2013omnidirectional} and magnonic logic devices \cite{rana2018voltage,khitun2010magnonic,ding2012realization,allwood2005magnetic}. The dispersion relation of the SWs is quadratic in the short wavelength limits and due to the Bragg scattering of SWs they form magnonic minibands which consist of allowed magnonic states with alternating forbidden band gaps. Apart from external magnetic field and material properties \cite{rana2012anisotropy,mahato2013configurational} the magnonic band structures can also be tuned by varying various physical and geometrical parameters of magnetic nanoelements and MCs such as shape  \cite{mahato2014tunable,saha2013time,mandal2013effects}, size  \cite{saha2015all,banerjee2014width,rana2011all}, arrangement of constituent nanoelements in MCs \cite{mahato2013configurational,rana2011detection}; lattice constant \cite{mandal2012optically} and lattice symmetry of MCs \cite{saha2013tunable}. Even magnonic bandgaps can be tuned effectively in bicomponent MCs. 

\begin{figure}[h!]
    \centering
    \includegraphics[width=1\linewidth]{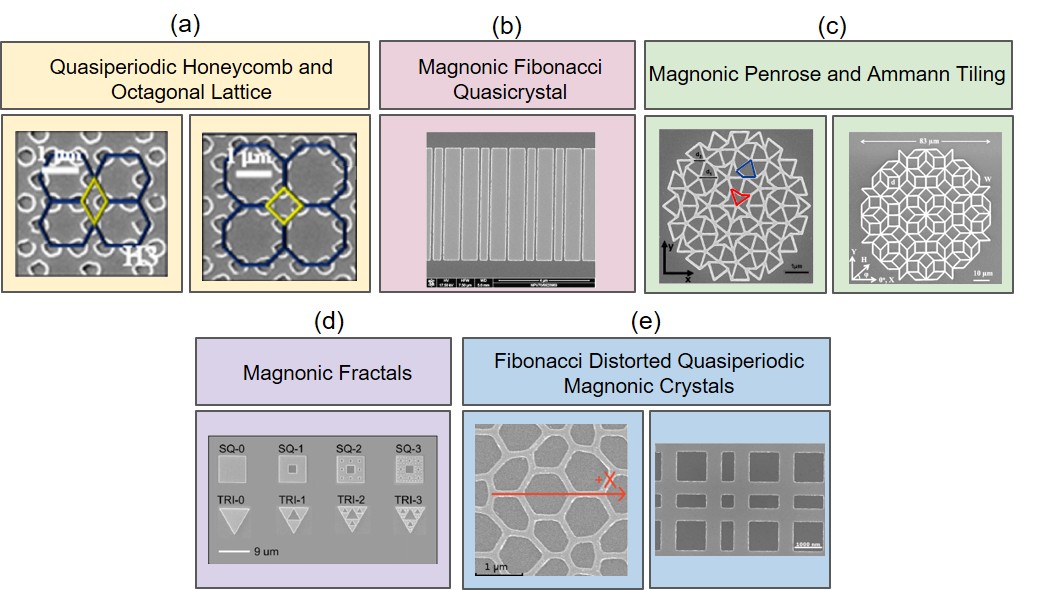}
    \caption{Schematic depicting quasiperiodic MCs. Scanning electron micrographs (SEM) of $(a)$ circular antidot arranged in quasiperiodic honeycomb and octagonal lattice, reproduced from \cite{majumder2021reconfigurable} $(b)$ Fibonacci sequences of magnetic nanostripes, reproduced from \cite{szulc2019remagnetization} $(c)$ Penrose $P2$ and Ammann tiling, reproduced from \cite{bhat2013controlled,bhat2014ferromagnetic} $(d)$ Sierpinski square and triangular structure, reprinted from \cite{zhou2022precessional} $(e)$ Fibonacci distorted quasiperiodic MCs, reprinted from \cite{frotanpour2020magnetization,farmer2015magnetic}.}
    \label{fig:Quasi_MC1}
\end{figure}

In conventional MCs the periodicity or translational symmetry is present, whereas in quasiperiodic MCs this translation symmetry is absent and have unconventional rotational symmetry which modifies the  SWs dynamics.  The recent advancement in nanolithography technology allows the fabrication of various unconventional nanostructures including 3D MCs and quasiperiodic MCs \cite{sebastian2020nanolithography}. Moreover, various sensitive experimental techniques like broadband vector network analyzer ferromagnetic resonance (VNA-FMR), Brillouin light scattering (BLS) spectroscopy and time-resolved magneto-optic Kerr effect (TRMOKE) microscopy have evolved with time so rapidly, that they allow us to probe the SWs from various complex magnetic nanostructures with very high frequency, wavevector, time and spatial resolutions \cite{rana2013magneto}.  Therefore, a significant amount of work has been done on quasiperiodic MCs like periodic MCs. The aperiodic arrangement of magnetic elements in quasiperiodic crystals gives rise to complex dispersion relations with irregularly spaced energy bands. This leads to novel phenomena such as multifractal behaviour and quasiperiodic localization of SWs modes \cite{costa2012fractal,rychly2018spin}. Notably, the density of states in periodic MCs exhibit regular patterns with sharp peaks corresponding to allowed SWs modes. In contrast, quasiperiodic MCs display intricate and irregular density of states with fractal characteristics, reflecting the complex arrangement of magnetic elements \cite{rychly2015spin}. Furthermore, the absence of translational symmetry in quasiperiodic MCs results in unique wave control properties. Quasiperiodic MCs exhibit enhanced wave localization and partial bandgaps, allowing for more efficient confinement of SWs within specific regions of the crystal. The confinement of SWs leads to potential applications of these crystals in magnonic waveguides and signal processing.

A number of reviews on the periodic MCs have already been devoted, conveying their characteristics, potential applications and various challenges \cite{krawczyk2014review,barman20212021,zakeri2020magnonic,barman2020magnetization}. However, to the best of our knowledge no review has been reported till date on quasiperiodic MCs. This review mainly focus on magnetization reversal dynamics and collective precessional dynamics of magnetization in quasiperiodic MCs. In this review article, we briefly discuss about different kinds of quasiperiodic MCs (figure \ref{fig:Quasi_MC1}) namely quasiperiodic hexagonal and octagonal lattice obtained by shifting the consecutive hexagonal and octagonal lattice, Fibonacci quasicrystal, magnonic Penrose and Ammann tiling, and magnonic fractals comprising of Sierpinski square, Sierpinski triangle. We also discuss the Fibonacci distorted quasiperiodic MCs. The review is concluded with various avenues of future exploration since the aperiodicity of these crystal lattices produces unrivalled tunability of magnetic band structure and stark modulation in the SWs dynamics, thereby unlocking their potential application in reconfigurable magnonic devices.

\section{Fabrication methods and measurement techniques}

The quasiperiodic MCs are usually fabricated through a combination of magnetron sputtering, electron beam evaporation, and electron beam lithography, followed by a liftoff process. To grow a thin film heterostructure over a substrate or resist pattern, magnetron sputtering and electron beam evaporation are usually employed. The subsequent application of electron beam lithography allows for precise pattering of the thin films. The details about the sample fabrication can be found in reference \cite{pimpin2012review}.  Further to prevent oxidation, a capping or protecting layer is usually deposited on the top of the magnetic films. The topography and roughness of the deposited samples are checked by using scanning electron microscope (SEM) and atomic force microscopy (AFM) techniques \cite{zhou2007fundamentals,giessibl2003advances}. The chemical composition and purity of the samples are measured using X-ray diffraction (XRD) and energy dispersive X-ray (EDX) techniques \cite{warren1990x,russ2013fundamentals}. The static magnetic properties of the samples are characterized by using vibrating sample magnetometry (VSM), superconducting quantum interference device (SQUID), and magneto-optic Kerr effect (MOKE) techniques \cite{rana2011magnetic,mahato2012magnetization,allwood2003magneto,barman2020magnetization}.

As discussed in the introduction section, the SWs dynamics of the quasiperiodic magnonic crystals can be measured through various techniques including TRMOKE, BLS and VNA-FMR. The magnetization dynamics in the time domain can be investigated using the TRMOKE technique which offers a time resolution of less than 100 femtoseconds, and it is used to probe the magnetization dynamics of a ferromagnetic material very efficiently from femto to nanosecond timescale \cite{barman2014time}. In this technique, a femtosecond laser pulse beam is used to excite or pump the electron, spin, and lattice systems within a  ferromagnet and the consequent magnetization dynamics are probes using another pulse laser beam after passing through a variable time delay by measuring the Kerr signal using a balanced photodetector, which completely isolates the Kerr signal from the reflectivity signals \cite{barman2007ultrafast,barman2008benchtop}.  The dispersion relation of the propagating SWs in a wavevector domain could be measured using a BLS technique where the light is inelastically scattered by the magnons present in the ferromagnetic system \cite{tacchi2012forbidden}. The spatial distribution of the SWs in magnetic nanostructures could also be imaged by using the micro-focused BLS technique which offers a spatial resolution of around 250 nm \cite{sebastian2015micro}.  
The magnetization dynamics of the magnetic system can also be measured in the frequency domain by using broadband ferromagnetic resonance spectrometer, which is based on a vector network analyzer \cite{kalarickal2006ferromagnetic}. Microwave signals with varying frequencies from the VNA are launched in the ferromagnetic system with the help of a coplanar waveguide by using a ground-signal-ground type of probe (GSG). At the resonance, the angular frequency of the microwave signal matches the precession frequency of magnetization in the material and an absorption in the signal is obtained. The real and imaginary parts of the scattering parameter in reflection (S11) and transmission geometry (S12) measured at various magnetic fields, give the the total SW spectra of the ferromagnetic system \cite{barman2018spin}.

\section{Quasiperiodic honeycomb and octagonal lattice}
\begin{figure}[!ht]
    \centering
    \includegraphics[width=1\linewidth]{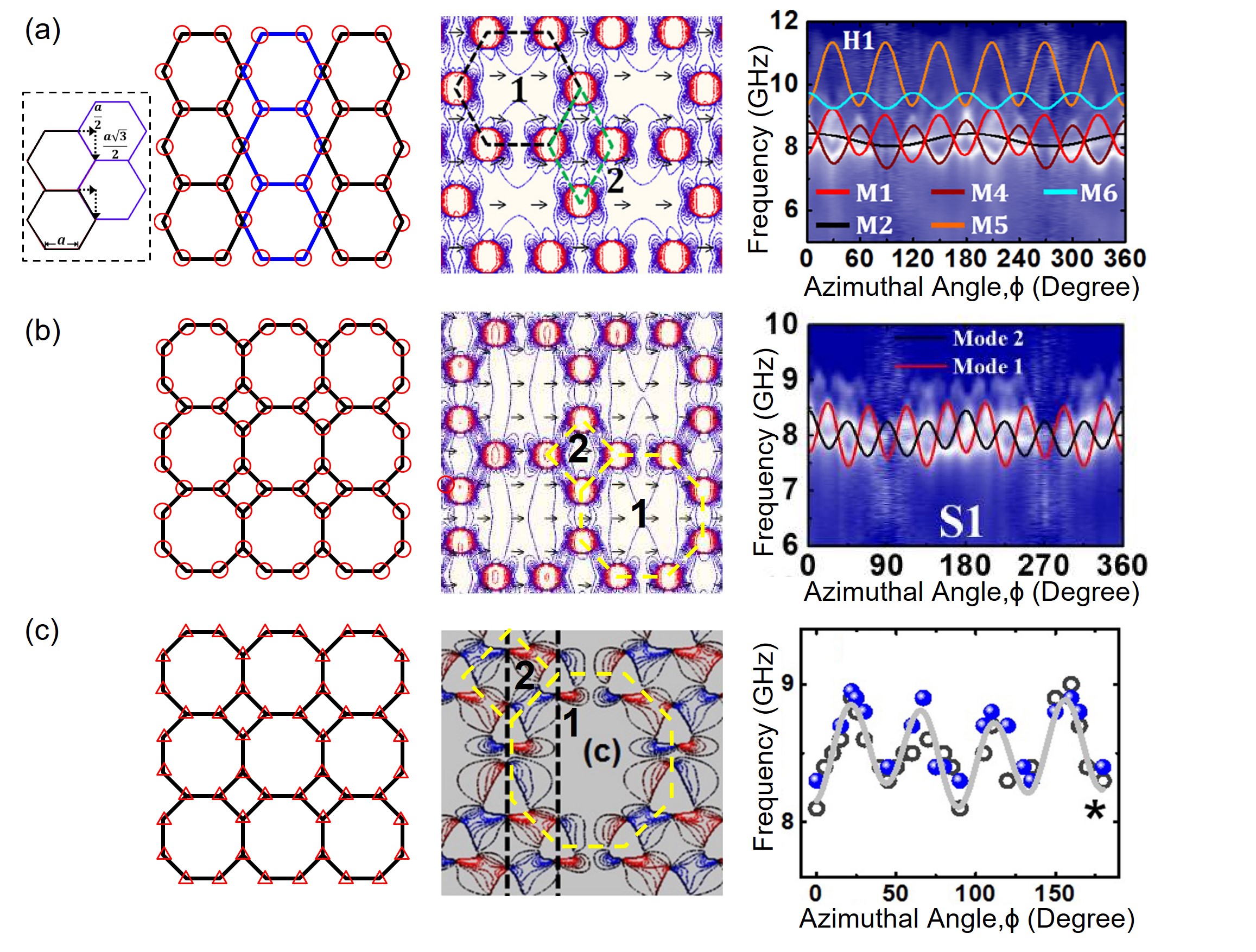}
    \caption{Left to right: figure showing the schematic, contour plots of the simulated magnetostatic field distributions and variation of SWs frequency as a function of the azimuthal angle for (a) quasiperiodic honeycomb antidot lattice obtained by shifting consecutive honeycomb cell, periodic honeycomb lattice is shown in the inset (b) quasiperiodic octagonal lattice with circular antidot (c) quasiperiodic octagonal lattice with triangular antidot. Reprinted with permission from \cite{choudhury2019controlled,choudhury2017efficient,de2021magnonic}. }
    \label{fig:QP_honey}
\end{figure}

Magnetization dynamics in the ferromagnetic dots (i.e., islands) and antidots (i.e., holes) arranged in various Bravais lattice configurations, including square, rectangular, and hexagonal arrangements have been studied extensively \cite{saha2013tunable,mandal2015tunable}. Bravais lattice is a periodic lattice exhibiting translation symmetry. The investigations have revealed that the SWs spectra exhibit notable variations owing to these lattice structures. SW dynamics in non-Bravais lattice like periodic honeycomb lattice has also been studied and due to the lack of full translation symmetry, it showed rich SWs dynamics as compared to the periodic lattices \cite{mondal2018influence}. 
Magnetization dynamics in the unconventional honeycomb antidot lattice with an artificial defect have been studied 
\cite{choudhury2019controlled}. As shown in figure \ref{fig:QP_honey}(a) artefact honeycomb lattice can be achieved by shifting the consecutive honeycomb lattice by $x=a/2$ and $y=(a\sqrt{3})/2$ with respect to the black honeycomb unit, having $a$ as the centre to centre distance of the consecutive lattice points. Due to the lack of translation symmetry, such a unique lattice is not a familiar Bravais lattice. This kind of lattice structure found similarity with the 2D stereographic projection of the extended rhombic dodecahedron of typical heusler alloy system \cite{shi2018prediction,rotjanapittayakul2018spin}. Rich SWs spectra were observed for most densely packed antidot lattice. Remarkably, as the periodicity of the lattice decreased, a systematic reduction in the number of SWs modes was unveiled arising due to the reduction of the demagnetizing field around the antidots. Ferromagnetic resonance (FMR) was used to measure the SWs spectra with the external magnetic field orientation, which disclosed the presence of anisotropic SWs modes with a superposition of six and two-fold rotation symmetry. Six-fold anisotropy arises due to the internal field inside the honeycomb cell whereas that within the rhombic region surrounding by honeycomb cells gives rise to two-fold anisotropy, shown by region 1 and 2 in figure \ref{fig:QP_honey}(a). 

In another study by Choudhury et al., they showed the magnetization dynamics of two-dimensional arrays of permalloy ($Ni_{80}Fe_{20}$) antidots with circular shape arranged in octagonal lattices which can be considered as quasiperiodic crystal due to the presence of the broken lattice symmetry \cite{choudhury2017efficient}. They observed that the SWs spectra exhibit a fascinating diversity, showcasing rich variations stemming from changes in the lattice constant, along with the strength and orientation of the applied bias magnetic field. The observed SWs frequency displays an eight-fold anisotropy characterized by the superposition of subtle four-fold anisotropy obtained within the rhombus and two-fold anisotropies that arise because of rectangular shape of the array as shown in figure \ref{fig:QP_honey}(b). Eight-fold and four-fold anisotropy arising from the angular diversity in the magnetostatic field distribution across specified regions of the octagonal lattice, shown by 1 and 2 has been depicted in figure \ref{fig:QP_honey}(b).

Like in the periodic MCs SWs dynamics can be varied by the size of the nanostructure \cite{saha2015all}. Majumder et al. studied the SWs dynamics in a 2D array of permalloy antidots of circular shape arranged in quasiperiodic octagonal and shifted honeycomb lattices by varying the diameter of the antidots with fixed lattice constant \cite{majumder2021reconfigurable}. The effect of strength and orientation of the in-plane external magnetic field on SWs dynamics is also investigated. A drastic modulation in the SWs spectra was attained with the variation in the antidot diameter due to the varying demagnetizing field. Moreover, higher demagnetizing effects of the rhombic shaped regions in the honeycomb lattice caused stronger localization of SWs modes as compared to the octagonal lattice.

Further, the SWs spectra in the octagonal lattice with the complex basis of the form of the triangle as shown in the figure \ref{fig:QP_honey}(c) has been reported by  De et al. \cite{de2021magnonic}. The quasiperiodic nature of octagonal lattice, combined with the presence of a complex triangular antidot basis lacking reflection symmetry, hugely modified the SWs dynamics. The magnetization dynamics by varying the orientation of the external magnetic field shows the presence of a prominent eight-fold anisotropy combined with a subtle three-fold anisotropy due to triangular antidot basis. SWs dynamics exhibit a systematic reduction of the number of modes as the lattice constant increases.  This fascinating behaviour is attributed because of the continuous reduction of the demagnetizing field around the antidots. Apart from this, the presence of localized edge mode was also observed in virtue of the sharp triangular corners of the antidot.

\section{Fibonacci quasiperiodic magnonic crystal}

\begin{figure}[!ht]
    \centering
    \includegraphics[width=1\linewidth]{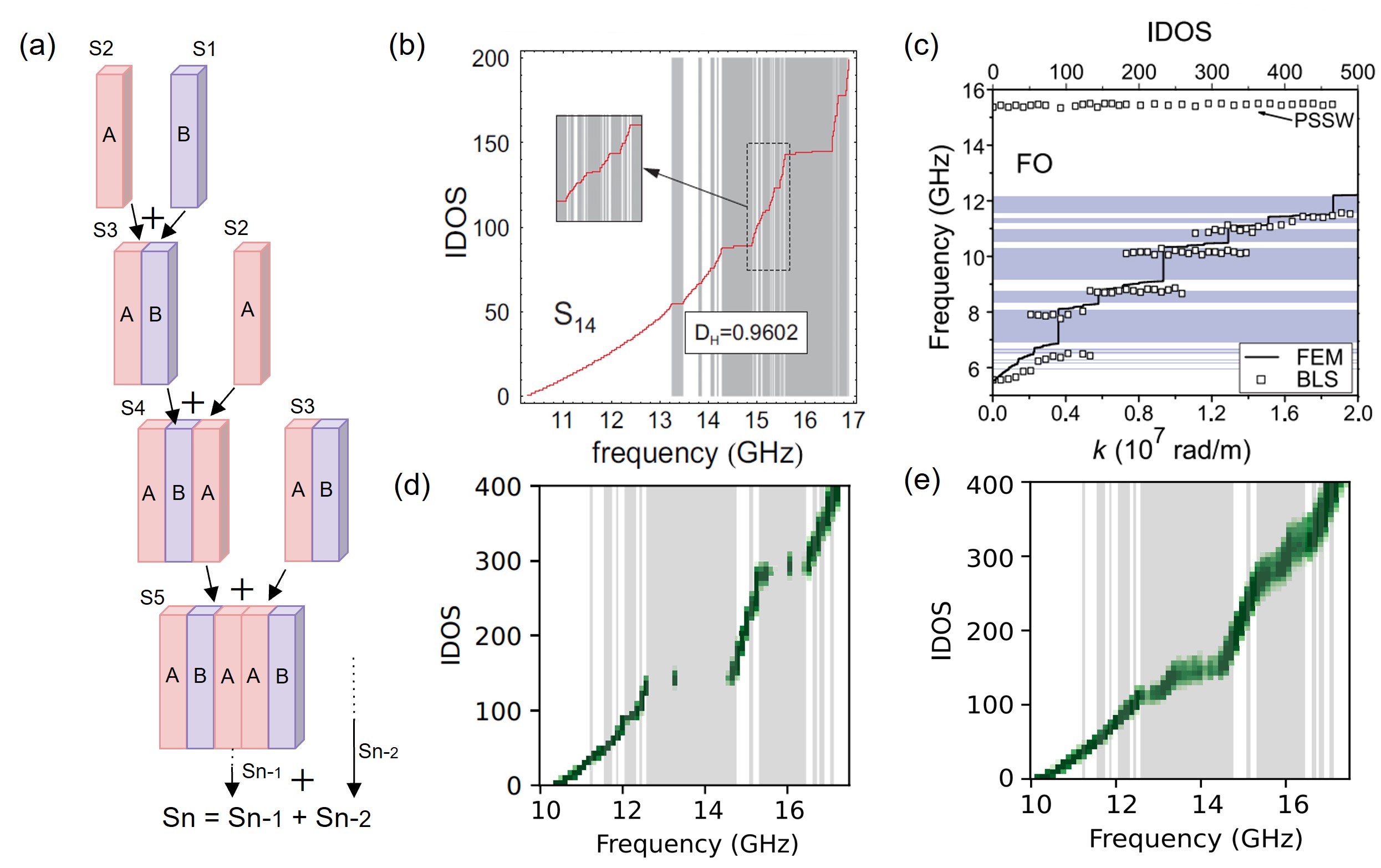}
    \caption{(a) Schematic showing the first few sequences of a planar Fibonacci quasiperiodic MC composed of strips having material A and B. (b) Calculated integrated density of states (IDOS) as a function of frequency, where grey area denotes the magnonic gap. Nestled within the IDOS plot lies the magnified region of the complex band gap structure, reprinted from \cite{rychly2015spin}. (c) Band structure obtained from BLS data and numerical calculation through finite element method (FEM) for ferromagnetic order, reprinted from \cite{lisiecki2019reprogrammability}. The IDOS as a function of frequency in a Fibonacci sequence of permalloy and cobalt stripes with incorporated defects has been presented, for phasonic defects of (d) 5\% and (e) 25\%. The grey area depicts the frequency gap in an ideal Fibonacci sequence with no phasonic defect. Reprinted with permission from \cite{mieszczak2022spin}.}
    \label{fig:QP_Fibonacci_m}
\end{figure}

The Fibonacci sequence is a sequence where the subsequent number is defined as the sum of its predecessors ($S_{n} = S_{n-1} + S_{n-2}$), giving rise to an ever-unfolding sequence which defies the traditional periodicity as shown in the figure \ref{fig:QP_Fibonacci_m}(a). This sequence appears in various natural phenomena, from the arrangement of leaves on plants to the spirals found in seashells and galaxies. Quasiperiodic crystals in the form of the Fibonacci sequence have already been studied for photonic \cite{vardeny2013optics,kohmoto1987localization}, phononic \cite{albuquerque2003theory} and electronic systems \cite{merlin1985quasiperiodic,laruelle1988fibonacci}. The Fibonacci sequence has also been implemented in the magnonic system to study the modification of magnetization dynamics \cite{coelho2011transmission,costa2012fractal,costa2013band}. SWs dynamics in nanostripes comprised of cobalt and permalloy arranged in the form of the Fibonacci sequence has been studied by Rychly et al. \cite{rychly2015spin}. The stripes were in direct contact ensuring the presence of both short range exchange and long range dipole interaction. The SWs spectra were analyzed by calculating the integrated density of states (IDOS) which is defined as:
\begin{equation}
     IDOS(f_i) = \sum_{j = 0}^{i} DOS(f_j).
\end{equation}

Where DOS is the density of SWs mode and $f_i$ is the frequency of $ith$ SWs mode. It has been found that the IDOS for quasiperiodic MCs consists of a complex, remarkable multilevel structure of frequency gaps as presented in the figure \ref{fig:QP_Fibonacci_m}(b). A magnified region of the spectra is also shown in the inset. The emerging multilevel structure of the magnonic gaps reveals the property of self-similarity, unlike the IDOS in periodic MCs which show a regular dependence on frequency. Further, the calculated non-integral value of the Hausdorff dimension of the spectrum shows the fractal property of the SWs spectra in quasiperiodic MCs. The Hausdorff dimension quantifies the self-similar or scaling nature of a set, providing a measure of its complexity derived from its scaling behaviour at different levels of magnification. A theoretical study on bicomponent magnonic crystal composed of cobalt and permalloy nanowires arranged in the form of Fibonacci sequence has also been investigated by Hussain et al. \cite{hussain2018quasiperiodic}. The Hamiltonian-based microscopic method was used to calculate the SWs spectra and showed the fractal-like scaling property of SWs in these systems. Grishin et al. have studied the magnetostatic surface SWs propagation in Fibonacci quasiperiodic MCs \cite{grishin2013dissipative}. They have observed that such quasiperiodic crystals have a large number of band gaps with narrow pass bands located between them. These feature of Fibonacci quasiperiodic magnonic crystal makes them suitable to be used in an active ring resonator for eigenmode selection. 

An experimental study on the SWs dispersion relation in a quasiperiodic MC has also been conducted using various techniques, including vector network analyzer - ferromagnetic resonance (VNA-FMR), Brillouin light scattering (BLS) spectroscopy, and scanning transmission X-ray microscopy (STXM) \cite{lisiecki2019reprogrammability}. Their investigation focused on permalloy nanowires with different widths, arranged in the Fibonacci sequence. They reported that changing the magnetization configuration of nanowires from ferromagnetic to antiferromagnetic order produces strong variation in SWs dynamics. Further, the hysteresis loops are measured through magneto-optic kerr effect (MOKE) and a wide plateau for antiferromagnetic order, allowing for a wider tuning range, are observed. The calculated IDOS through both BLS and finite element method (FEM) showed several narrow and well-resolved peaks for both ferromagnetic and antiferromagnetic order as depicted in figure \ref{fig:QP_Fibonacci_m}(c). This dispersion character is quite different from the periodically arranged nanowires of different widths where the mode follows the Bloch theorem, revealing a continuous dispersion and periodic oscillation of frequencies.

Another study by Szulc et al. is focused on the magnetization reversal process in one-dimensional magnetic structure comprised of permalloy nanostripes of varying lengths, meticulously arranged with air gaps in adherence to the Fibonacci sequence \cite{szulc2019remagnetization}. Magnetization reversal was measured by MOKE to compare the hysteresis loops between the permalloy nanostripes with periodic arrangement and arranged in a Fibonacci pattern. An observable variation in the magnetization at the plateau was detected, suggesting its origin in response to distinct stray magnetic fields. Furthermore, modifying the dimensions of the nanostripes, in terms of both thickness and length, induces a significant modulation in magnetostatic interactions, thereby in the magnetization reversal dynamics.

The lifetime of the SWs eigenmode in permalloy and cobalt nanowires arranged in a periodic manner and in a Fibonacci sequence has also been studied \cite{rychly2016spin}. The material is chosen in such a way that one (i.e. permalloy) has lower saturation magnetization and lower damping while the other one (i.e. cobalt) has higher saturation magnetization and higher damping parameters. It is observed that for these systems the lifetime of eigenmodes can be prolonged beyond that of modes within a uniform slab crafted from material that possesses parameters derived from volume averaging of parameters from both cobalt and permalloy elements. They observed a correlation between spatial distribution of SWs amplitudes and their lifetimes. For the periodic system, the monotonic increase in the lifetime within the individual band has been observed whereas a jump in lifetime can be seen in the quasiperiodic system, which indicates that damping in magnonic nanostructure can be control by introducing periodic as well as quasiperiodic modulation.

Propagation of SWs in Fibonacci quasiperiodic crystal consisting of dipolar coupled nanowires of different sizes has also been experimentally investigated \cite{lisiecki2019magnons}. The time-resolved scanning transmission X-ray microscope was used to image the SWs. They demonstrate that the propagation of SWs transcends the confinement of the long-wavelength limit, and the structure can be considered as an effective medium. This phenomenon extends to higher frequencies, where the profound long-range quasiperiodic order of the structure emerges as a crucial factor governing SWs propagation.  Additionally, the experimental observation revealed the mini band gaps and the SWs localization. These intriguing phenomena directly emerge as a result of the collective effects of SWs within quasiperiodic MCs.

When structural disorders are carefully introduced in the Fibonacci structure, the SWs spectra are significantly modified. Mieszczak et al. in their study introduced several stages of disformation known as phasonic defects, where the disorder is introduced in the system through rearrangements of the permalloy and cobalt stripes \cite{mieszczak2022spin}. On moving from quasiperiodic order to increasing disorder, they observe a gradual degradation of SW fractal spectra and closure of frequency gaps as shown in the figure \ref{fig:QP_Fibonacci_m}(d) and (e). The introduction of phasonic defects caused the vanishing of Van Hove singularities which occur when there is a sharp peak or divergence in the DOS at a specific energy. The phasonic defect gives rise to new modes in the system at the edge of the frequency gap. As disorder levels escalate, these newly formed modes disperse, eventually leading to the closure of the gaps as depicted in figure \ref{fig:QP_Fibonacci_m}(e). The study revealed that in the magnonic system having both short-range exchange interaction and long-range dipole interaction, the phasonic defect causes the closing of small gaps and enhancement of mode localization. This analysis can be further extended to study the effect of phasonic defects within two-dimensional magnonic systems leading to their fascinating application potential in magnonic signal processing.

\section{Quasiperiodic magnonic crystal in the form of Penrose and Ammann tiling}
Penrose and Ammann tilings, alluring creations at the intersection of art and mathematics, have attracted minds with their beautiful patterns that defy traditional concepts of translational symmetry and repetition. These complex arrangements, discovered by mathematicians Sir Roger Penrose and Robert Ammann, challenged the perceptions of periodicity. 
While these structures have already been studied in the field of photonics \cite{vardeny2013optics,kaliteevski2000two,ma2022quasiperiodic} and phononics crystals\cite{chen2015band,han2020super}, a new horizon beckons. Our journey now takes us into the territory of quasiperiodic MCs. Thus, we embark on an odyssey to uncover the connections between Penrose, Ammann, and the tantalizing potential of quasiperiodic magnonic wonders.

Magnetization reversal and dynamics in quasiperiodic antidot lattice in the form of the Penrose tiling have been studied by Bhat et al. \cite{bhat2013controlled}. Penrose $P2$ antidot tiling having \enquote{kites} and \enquote{darts} (shown by red and blue colours in figure \ref{fig:Quasi_MC1}(c)) as the antidot with permalloy boundaries has been fabricated over the silicon substrate with electron beam lithography as shown in the figure \ref{fig:Quasi_MC1}(c). They have observed two contrasting SWs modes based upon the magnitude of the applied DC field: within saturation region, the \enquote{symmetric} SWs mode traits mirror themselves symmetrically on either side of the field origin. Conversely, the \enquote{asymmetric} SWs modes, exclusively confined to the lower-field domain, are characterized by their appearance only on one side of the field origin during a field sweep from -6 KOe to 6 KOe. This suggests that the consistent patterns of modes observed during changes in the magnetic field are likely due to the magnetization being pinned parallel to the antidot edges and the confinement of domain wall at the vertices of Penrose tiling is responsible for asymmetric SWs mode at low-field region. Additionally, the SWs spectra of the Penrose tiling exhibit ten-fold rotation symmetry with the azimuthal orientation of the external magnetic field. The pinning of magnetization along segment edges and the confinement of domain walls at vertices within quasiperiodic lattices are responsible for a significant influence over the SWs dynamics. This phenomenon underscores the emergence of Penrose $P2$  as a fresh paradigm within the domain of quasiperiodic MCs.

The SWs dynamics of quasiperiodic MC in the form of Ammann tiling have been measured using a broadband FMR and the presence of an eight-fold rotational symmetry was observed \cite{bhat2014ferromagnetic}. Ammann antidot tiling having \enquote{square} and \enquote{rhombi} as the antidot with permalloy boundaries were patterned over the silicon substrate using the electron beam lithography as shown in the figure \ref{fig:Quasi_MC1}(c). A rich/multimode SWs spectra with compared to the SWs spectra for a periodic antidot lattice has been observed, which arises due to the complex topology and lower symmetry. Furthermore, in the proximity of saturation, the FMR experiments revealed the symmetrically positioned SWs modes, while in the hysteretic regime, the distinctive asymmetric SWs modes were observed, similar to the finding of Penrose $P2$ tiling.

Rychly et al. investigated the propagation of the SWs in the 2D bicomponent quasiperiodic MCs in the form of Penrose $P3$ coverage \cite{rychly2018spin}. Penrose $P3$ tiling consists of two rhombi tiles having different acute angles. In order to manifest the long-range order arrangement, a disk of a different kind of ferromagnetic material is included in the centre of every rhombus. Unlike previous studies that employed a small volume fraction of magnetic material in the form of magnetic wires which can be characterized by high anisotropy due to the large antidot shapes, this study minimizes the impact of shape anisotropy. The computed IDOS revealed distinct plateaus within the spectra, indicating frequency gaps that signify the long-range order present within the quasiperiodic MC. Moreover, these gaps were clearly visible in structures where there were significant differences in the properties of magnetic materials used, and when there was a large filling fraction, often associated with large embedded disks.

 \begin{figure}[!ht]
    \centering
    \includegraphics[width=1\linewidth]{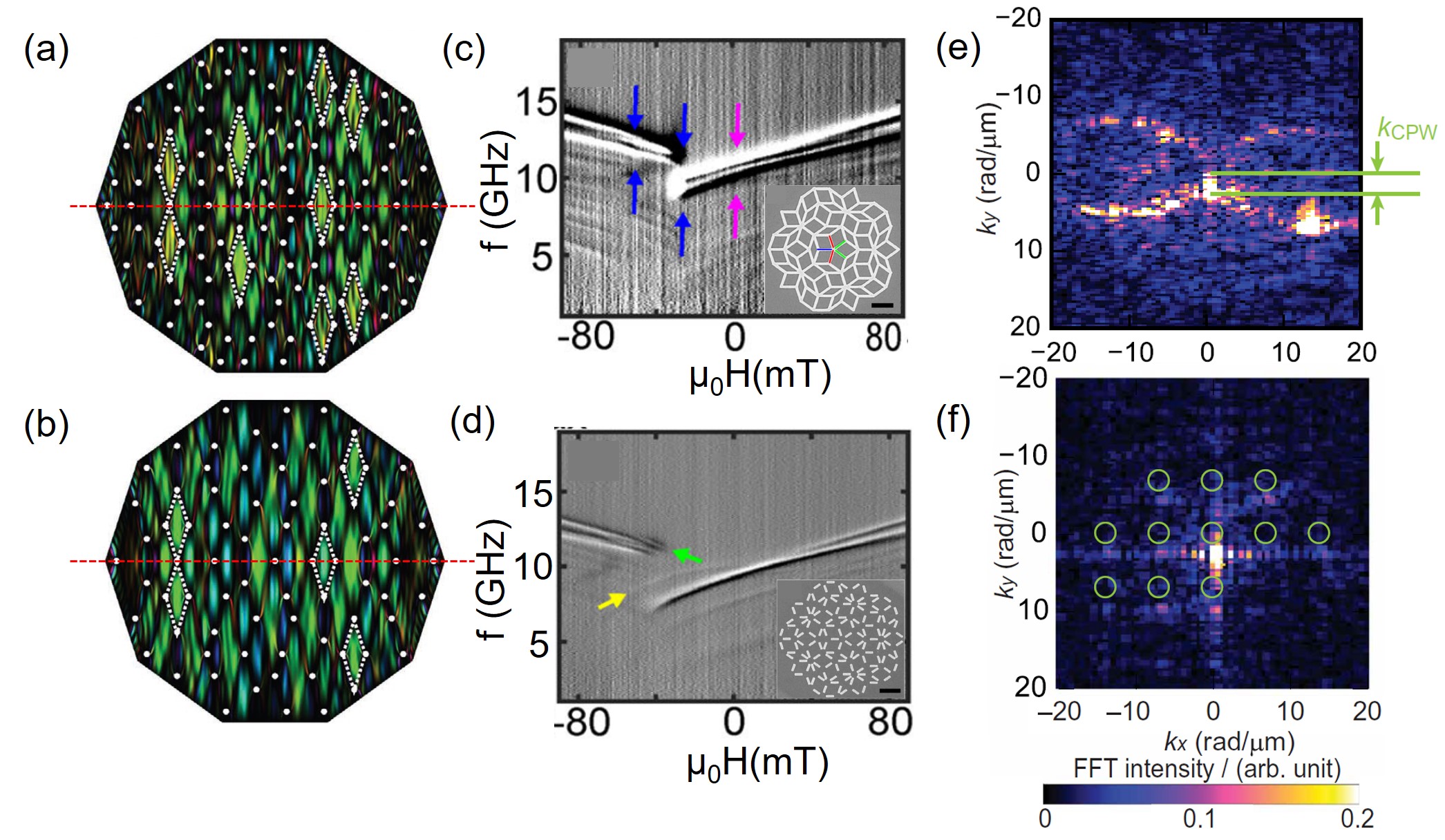}
    \caption{Maps of SWs precessional motion for Penrose (a) $P2$ and (b) $P3$ quasiperiodic nanohole lattice, white dashed lines highlight spin precession surrounded by four nanoholes (rhombi), reprinted from \cite{watanabe2020direct}. Broadband SWs spectra for Penrose $P3$ tiling where (c) nanobars are fully connected, (d) nanobars are disconnected, reprinted from \cite{bhat2023spin}. Fourier transform of SWs in (e) Penrose $P3$ tiling (f) Periodic MC, reprinted with permission from \cite{watanabe2021direct}.}
    \label{fig:Ammann_QC_h}
\end{figure}

SWs spectroscopy in Penrose tiling using the spatially resolved Brillouin light scattering (BLS) has also been studied by Watanabe \cite{watanabe2020direct}. The nanohole lattice was prepared by etching out the circular holes on the vertices of Penrose $P2$ and $P3$ quasiperiodic tiling. As proposed earlier, a ten-fold angular symmetry in the SWs modes with the variation of the azimuthal angle of the external magnetic field has been observed. The aperiodic nanohole patterns give rise to stripe-like excited modes. These magnon nanochannels incorporate aperiodic sequences of bends, and this aperiodicity give rise to varying magnon band structures across channels, setting them apart from their periodic counterparts. It is noted the SWs spectra indicate the emergence of worm-like nanochannel and magnonic motif (here white dotted region) shown in figure \ref{fig:Ammann_QC_h}(a) and (b). Moreover, these SWs modes display mirror symmetry (here about the red dotted line in figure \ref{fig:Ammann_QC_h}(a) and (b)) and the axis of mirror symmetry changes along with the direction of the applied magnetic field. The identification of worm-like nanochannels within 2D antidot quasicrystals resulted in an unparalleled demultiplexing mechanism involving microwaves.

A recent study has been performed on Penrose and Ammann tiling with different exchange and dipolar interactions \cite{bhat2023spin}. Quasiperiodic MC in the form of Penrose and Ammann tiles has been created by permalloy nanobars. The interconnected nanobars joined at the vertices have both exchange and dipolar coupling whereas partially connected nanobars are only dipolarly coupled as shown in the inset of figure \ref{fig:Ammann_QC_h}(c) and (d). It has been observed that, for interconnected nanobars shown in figure \ref{fig:Ammann_QC_h}(c), the two prominent branches appeared at larger frequencies compared to the disconnected nanobars shown in figure \ref{fig:Ammann_QC_h}(d). The length of the nanobars is shorter for figure \ref{fig:Ammann_QC_h}(d) with respect to figure \ref{fig:Ammann_QC_h}(c), which leads to a reduction of the total internal fields subsequently leading to a reduction in resonance frequencies. Additionally, due to the asymmetric and aperiodic patterns around each vertex, nanobars oriented at the same angle to the applied magnetic field exhibit distinct switching behaviours which is influenced by their specific local surroundings. In another study, systematic variations and consistent patterns of field-dependent resonance frequencies were observed for Penrose and Ammann tiling structures having interconnected magnetic nanobar \cite{bhat2018angle}. Through a comprehensive analysis of experimental and simulated data within the saturation range, it becomes evident that the resonance frequency is primarily governed by the shape anisotropy of individual nanobars especially when accounting for the impact of various long range ordered local environments under external field conditions.

The SWs dynamics in Penrose $P3$ tiling having nanotroughs etched out at the vertices on the low damping ferrimagnetic material (yttrium iron garnet), has also been documented \cite{watanabe2021direct}. The wave vector of the SWs for square etched periodic MC and quasiperiodic MC in the form of Penrose $P3$ tiling has been determined. The SWs map in quasiperiodic lattice revealed the presence of irregular wavefronts, shown in figure \ref{fig:Ammann_QC_h}(e), in contrast to those observed in periodic magnonic lattice, shown in figure \ref{fig:Ammann_QC_h}(f). For Penrose tiling the frequency contour resembling a dumbbell shape signifies the propagation of dipole-dominated SWs in an omnidirectional manner as shown in the figure \ref{fig:Ammann_QC_h}(e), which is attributed to the unconventional rotational symmetry in the quasiperiodic Penrose lattice. This unconventional symmetry offers benefits to integrated magnonic circuits, by allowing efficient multidirectional magnon emission at a single frequency. Additionally, the presence of bandgaps offers the potential for magnon waveguides without the need for a strong out of plane magnetic field.

\section{Quasiperiodic magnonic fractals}
Fractals, with their mesmerizing blend of intricate aesthetics and mathematical complexity, seamlessly connect the realms of art and science. These mysterious patterns defy traditional notions of symmetry and repetition, sparking inspiration and discoveries across various disciplines \cite{fremling2020existence,van2016quantum}. As they've woven their influence into the tapestry of art and mathematics \cite{falconer2004fractal}, fractals have also made a lasting impression in the worlds of photonics \cite{yang2020photonic} and plasmonics \cite{de2018multiband}. Moreover, Hofstadter's butterfly fractal patterns, named after physicist Douglas R. Hofstadter, exemplify the profound connections between mathematics and quantum physics. These intricate butterfly patterns reveal the interplay of electrons in a magnetic field, offering insights into the quantum Hall effect and serving as a bridge between the abstract world of physics and the aesthetic domain of visual art.

\begin{figure}[!ht]
    \centering
    \includegraphics[width=1\linewidth]{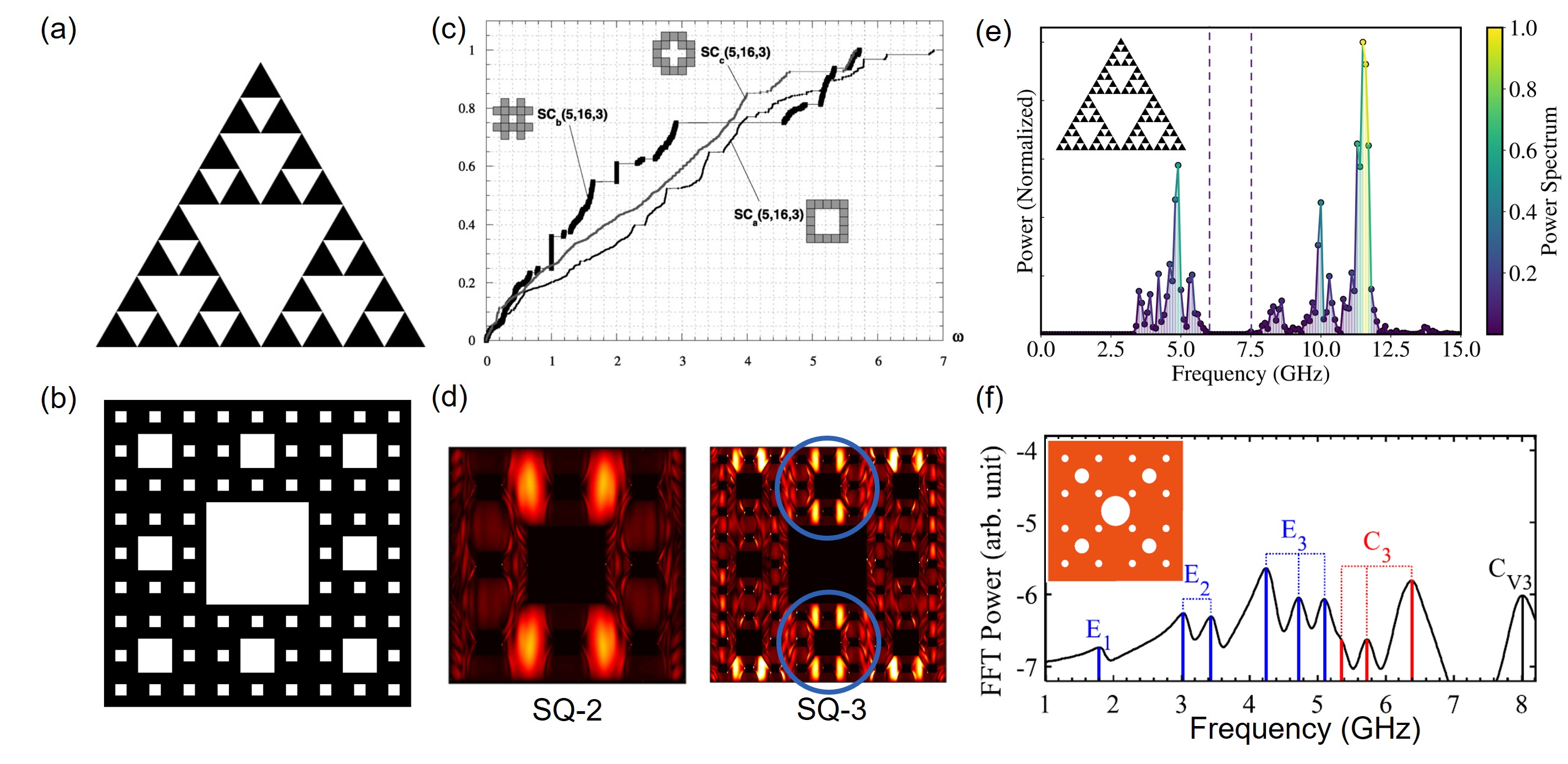}
    \caption{Schematic of (a) Sierpinski gasket and (b) Seirpinski Carpet. (c) The normalized integrated density of state for the Sierpinski carpet having different fractal connectivity but has a common fractal dimension of 1.723, reprinted from \cite{monceau2010spin}. (d) Simulated distribution of the amplitude of the SWs mode for the second and third iteration of Sierpinski carpet i.e. SQ-2 and SQ-3 having a bias field of 15 mT, reprinted from \cite{zhou2022precessional}. (e) FFT spectra derived from simulated time-domain magnetization data of Sierpinski triangle for the 4th iteration under an external magnetic field ($H_{ext}$) of 0.1 T and at an azimuthal angle $\phi = 0^{\circ}$, where the region between the purple dotted lines shows the frequency gap, adapted from \cite{mehta2023tunability}. (f) SWs mode spectra in antidot lattice fractal having third iteration. Where $E_3$ and $C_3$ denote the edge and centre mode, having a biased magnetic field of 30 mT in the +x direction, reprinted with permission from \cite{park2021recursive}.}
    \label{fig:fractal}
\end{figure}

Fractals can be assigned into two different groups: random and deterministic fractals. Both these structures exhibit, by definition, a noninteger Hausdorff dimension and can be characterized by self-similarity and scale invariance. The prevalence of random fractals in nature, seen in Romanesque cauliflower, broccoli, snowflakes, ferns, frost, and thunderstorm lightning patterns, is widespread, where the self-similarity is randomly distributed. In contrast, deterministic fractals offer a unique form of self-similarity that defies random distribution. At every level of magnification, a portion of the structure recapitulates the entirety of the whole. Sierpinski gasket shown in figure \ref{fig:fractal}(a) and Sierpinski carpet shown in figure \ref{fig:fractal}(b), named here as SC(n,p,i), share common origins in systematic iteration. To generate a Sierpinski carpet, initially, a square is divided into $n^2$ equal subsquares, with $n^2 - p$ subsquares are removed. This process is repeated i times, resulting in SC(n,p,i) composed of $n^i$ self-similar pieces, of which $p^i$ are occupied. The fractal dimension known as the Hausdorff dimension, is defined by $d = \ln{(p)} / \ln{(n)}$, encapsulates the relationship between p (occupied subsquares) and n (magnification factor).

Now, as we stand on the precipice of exploring magnonic fractals, we embark on a journey to unravel the compelling links between these mathematical marvels and the transformative possibilities they hold in the realm of magnonics \cite{barnsley2014fractals,kigami2001analysis,feder2013fractals}. Monceau et al. theoretically studied the SWs spectra of deterministic Siepinski carpets \cite{monceau2010spin}. The calculated integrated density of states of magnetic excitations, reveals a set of spectra characterized by a staircase-like pattern, which are singular continuous functions of the frequency, featuring numerous gaps and plateaus, depicted in the figure \ref{fig:fractal}(c). The SWs spectra demonstrate their sensitivity not only to the fractal dimension but also to the connectivity properties which refers to how the different parts of the fractal are connected as shown in figure \ref{fig:fractal}(c). This finding is closely associated with the emergence of fractal subdimensions derived from the eigenvalues of the connectivity matrix governing the formation of the fractal structure. In summary, this study reveals a prominent trait inherent to deterministic fractals: the association between the integrated density of states IDOS and fractal subdimensions. Remarkably, the presence of connectivity emerges as a reliable indicator for mode localization. In another work by Nowak et al., they explored the phase diagram of diluted Ising antiferromagnets under the influence of high external magnetic fields \cite{nowak1991diluted}. Within this investigation, they identified the emergence of a spin glass phase characterized by the presence of a stable domain state.  Upon analyzing these domains, they found that these domains exhibit fractal-like structures.

Experimental studies have also been carried out to explore the SWs dynamics in the magnetic fractal structure. Swoboda et al. studied the SW spectra of permalloy Sierpinski carpets by means of broadband FMR measurements and micromagnetic simulations \cite{swoboda2015control}. Sierpinski carpet having dimensions 1.893, 1.792 and 1.723 has been generated using the iteration process as described above. In all the cases it was observed that as iteration number advanced there was an increase in non-uniformity in the internal field of the carpets. Consequently, this led to a noticeable escalation in the number of quantized SW modes. Regardless of the geometric parameters, the number and frequencies of SW modes within Sierpinski carpets can be controlled by adjusting the external magnetic field angle, which modifies the demagnetization field, consequently affecting the frequencies and complexity of the SW modes. Furthermore, the simulated spatial profile shows the localization of SW modes depicting the quasiperiodicity.

Magnetization reversal in the permalloy Sierpinski triangle has been studied by Dai et al. \cite{dai2019controlled}. The Sierpinski triangle has been generated using a scaling factor of $1/2$, where the equilateral triangle has been used as a generator. The subsequent action involves eliminating a central triangle, with its vertices positioned at the halfway points along the edges of the initial triangle. This procedure is iterated for the remaining three triangles, resulting in the creation of Sierpinski triangle structures. It has been found that as the iteration number increases, the coercivity and remanence ratio given by $M/M_s$ of the subsequent iterations grows due to the reduction in triangle size. However, in comparison to a single triangle of the same size, the fractal structure exhibits significantly lower coercivity. This is due to the stray magnetic field generated by adjacent triangles within the fractal structure exerting a notable influence on the fractal's magnetization reversal process. Moreover, to understand the effect of fractal structure on the magnetization, the corresponding Barkhausen noise signal which is defined as the ratio of change in magnetization with the change in applied field has also been studied. Numerous jumps have been observed in magnetic hysteresis for a particular iteration number and the number of jumps increases with the increase of the iteration numbers. That reveals the gradual broadening of the signals and a decrease in peak intensity as the iteration number rises, which suggests that magnetization within individual triangles of the Sierpinski triangle becomes increasingly disordered due to structurization with the iteration numbers. 

In another work by Zhou et al., the magnetostatic SWs mode in Sierpinski square and triangle has been imaged by using a time-resolved scanning kerr microscope (TRSKM) \cite{zhou2022precessional}. In order to generate Sierpinski square a scaling factor of $1/3$ has been used. The precessional dynamics were examined in samples featuring a progression from basic geometric patterns to more complex Sierpinski fractals. In the case of SQ-3 which is referred to as the third iteration of Sierpinski carpet, a connection between its simulated dynamics and the modes present in SQ-2 (second iteration) through a scaling relationship has been observed.  As shown in figure \ref{fig:fractal}(d) the distribution of amplitude of SWs mode in SQ-3 (region marked by the blue circles) is similar to the distribution of amplitude of SWs mode in SQ-2. It becomes evident that the distribution of magnetostatic mode amplitudes follows the geometric scaling principles. That gives rise to the formation of scaled mode patterns within fractal structures, but with an additional iteration. However, achieving this requires careful consideration of the magnetic boundaries and the exclusion of regions where edge modes are present. Additionally, unlike the dominant precessional dynamics observed in periodic MCs, which primarily rely on unit cell eigenmodes and translation symmetry, magnetic fractals present a more complex amplitude distribution. This distribution is governed by geometric scaling and imitates the characteristics of geometric structures across various length scales.

In a recent work by Mehta et al., SWs dynamics and its tunability have been observed in triangular-shaped magnetic fractals \cite{mehta2023tunability}. Permalloy Sierpinski triangle having a geometrical scaling factor of $1/2$ has been simulated to understand the magnetization dynamics and their application. It has been observed that SWs dynamics change significantly with the iteration number with the appearance of a frequency gap with an iteration number exceeding some certain value as shown in figure \ref{fig:fractal}(e). SWs dynamics in individual triangular building blocks have also been compared and are quite different from the Sierpinski triangle. The phase distribution of SWs modes reveals that within a larger triangle, the phase experiences continuous changes throughout the triangle giving rise to intra-triangular magnetostatic interactions. In contrast, within a smaller triangle, a uniform phase distribution is observed due to the convergence of a multi-domain system into a single-domain, resulting in inter-triangular magnetostatic interactions. SWs dynamics in the periodic triangular array have also been calculated. The dynamics exhibit significant distinctions when compared to those of a Sierpinski triangle, yet the frequency gaps are present in both scenarios. Furthermore, the spatial profile of SWs modes within the Sierpinski triangle demonstrates the localization of spin modes, a distinct hallmark of fractal behaviour. Additionally, it has been noted that the tuning of SWs spectra is achievable through variations in the strength and orientation of the external magnetic field. A six-fold symmetry is observed for such system. The appearance of the frequency gap and the tunability of SWs spectra depicts the potential application of the Sierpinski triangle as a nanoscale field controlled magnonic device, like SWs filters or SWs splitters.

SWs dynamics in ferromagnetic antidot fractal lattice with periodic boundary conditions has also been investigated by Park et al. \cite{park2021recursive}. The concept behind modelling fractal antidot lattices involved creating a series of antidot lattices that share self-replicating characteristics, all within an integer Hausdorff dimension framework. The antidot lattices exhibit a self-similar pattern in their geometric parameters, specifically in terms of diameter ($D$) and lattice constant ($L$) and the 2D periodic lattice used here has a square Bravais symmetry. For each nth antidot lattice ($A_n$), the values of $D_n$ and $L_n$ are precisely half of those of the preceding lattice, $A_{n−1}$. To illustrate, consider the progression from $A_1$ to $A_4$: $A_2$ features $L_2$ = $L_1/2$ and $D_2$ = $D_1/2$, and this pattern continues. Next, the nth fractal ($S_n$) is formed by superimposing the individual antidot lattices, starting with $S_1$ = $A_1$ and subsequently adding $A_2$, $A_3$, and so forth, resulting in $S_n = A_1 + A_2 + A_3 + ... + A_n$, $S_3$ has been shown in the inset of \ref{fig:fractal}(f). It has been found that the SWs eigenmode within the antidot lattice fractal undergoes a splitting into multiplets and this can be attributed due to the fractals’ inhomogeneous and asymmetric internal magnetic fields. Furthermore, it has been observed that the recursive development of geometrical fractals gives rise to the same recursive evolution of SWs multiplets, SWs mode spectra for $S_3$ have been shown in the figure \ref{fig:fractal}(f).

\section{Fibonacci-distorted quasiperiodic magnonic crystals}

\begin{figure}[!ht]
    \centering
    \includegraphics[width=0.8\linewidth]{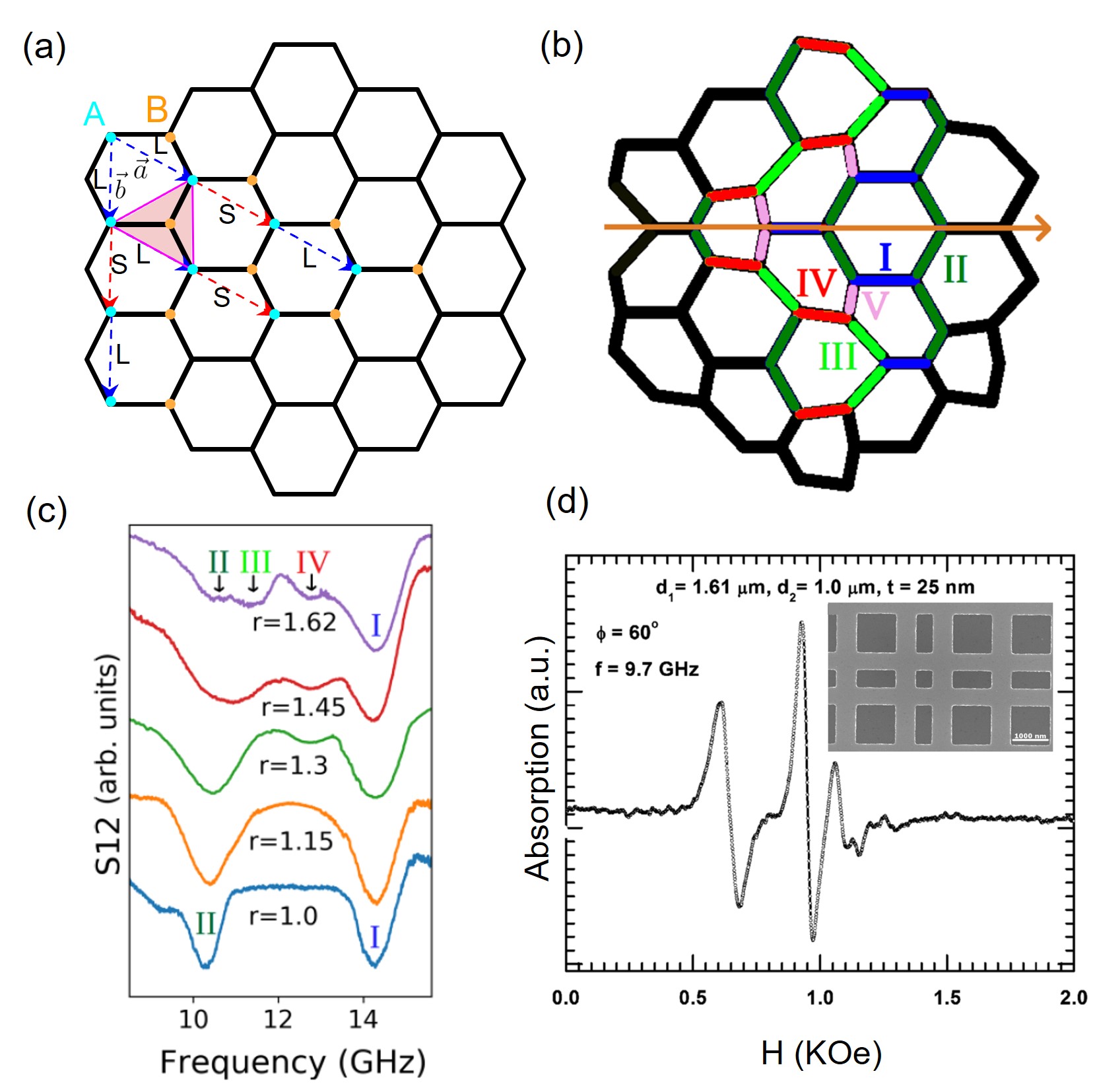}
    \caption{ (a) Undistorted kagome artificial spin ice having $a$ and $b$ as the primitive translation vectors. In the distorted lattice, the primitive vectors lengths follow the Fibonacci sequence, with \enquote{L} represented by a long length in blue and \enquote{S} represented by a short length in red. (b) Depiction of the distorted kagome artificial spin ice structure is provided for a specific ratio, $r$ = L/S = 1.62. (c) Ferromagnetic resonance (FMR) outcomes under an applied magnetic field H = 1000 Oe, varying with different distortion ratios $r$, reprinted with permission from \cite{frotanpour2020magnetization}. (d) FMR absorption spectra against the applied field for Fibonacci distorted square antidot lattice. The DC field was oriented at an angle $\theta$ of 60 degrees, inset showing the SEM image of Fibonacci distorted square antidot lattice having long and short lattice spacings, reprinted from \cite{farmer2015magnetic}.}
    \label{fig:fib_dis}
\end{figure}

As previously defined, the 1D Fibonacci lattice exhibits a long-range ordered arrangement, effectively representing a continuous deformation of a periodic lattice. In this context, the quasiperiodic Fibonacci sequence serves as a tool for distorting periodic crystals and generating quasiperiodic structures, thereby providing avenues for precisely tuning the dynamics of SWs. This Fibonacci pattern is also evident in Penrose tiling, where a Penrose P2 tiling showcases a distinctive Fibonacci arrangement of long \enquote{L} and short \enquote{S} spacings within the planes of parallel segments, with the flexibility of deriving this sequence from any of its five mirror planes. Frotanpour et al. studied the magnetization dynamics of a permalloy kagome artificial spin ice structure subjected to Fibonacci distortion \cite{frotanpour2020magnetization}. Artificial spin ices, which are periodic lattices suppressing long-range magnetic order due to their frustrated topology, consist of interconnected nanomagnets organized on diverse lattices \cite{skjaervo2020advances,nisoli2013colloquium}. These metamaterials exhibit interesting phenomena such as the emergence of magnetic monopoles \cite{ladak2010direct}, phase transitions \cite{levis2013thermal} and collective dynamics \cite{mamica2018spin}. Additionally, symmetry breaking in artificial spin ice leads to the tunability of SWs spectra opening their way as reconfigurable MCs \cite{saha2021spin,gliga2020dynamics,krawczyk2014review}. Kagome artificial spin ice, in particular, features a honeycomb lattice structure. The distorted arrangement can be achieved by replacing lattice translation vectors $a$ and $b$ of honeycomb lattice with a Fibonacci pattern of long \enquote{L} and short \enquote{S} distances as shown in the figure \ref{fig:fib_dis}(a). The ratio ($r$) of L to S can vary from 1.00 (undistorted), as depicted in figure \ref{fig:fib_dis}(a) to 1.62 (highly distorted), illustrated in figure \ref{fig:fib_dis}(b). With this distortion, the honeycomb lattice with sixfold rotation as well as mirror symmetry has been reduced to a quasiperiodic crystal having only mirror symmetry shown by the orange arrow in figure \ref{fig:fib_dis}(b).  Through experiment and micromagnetic simulation, it has been observed that the Fibonacci distortion leads to the widening and even division of ferromagnetic resonance (FMR) modes into multiple branches. Figure \ref{fig:fib_dis}(c) provides a clear visual representation of the frequency shifts, broadening of ferromagnetic resonance (FMR) modes, and the emergence of new FMR modes as the ratio $r$ increases. Further, it has been observed that the reversal dynamics and the precise characteristics of FMR modes, including their well-defined frequencies and frequency-field slopes, can be tuned by altering the extent of lattice distortion. This tunability of resonance frequencies and bandwidth depicts the potential application of Fibonacci distorted kagome artificial spin ice in magnonic devices. Moreover, a recent study by Giovannini et al. introduced the dynamical matrix method to calculate the magnetic normal modes of the magnonic system \cite{giovannini2021magnetic}. This method offered a theoretical framework to understand the frequency of SWs modes in Fibonacci-distorted artificial spin systems by addressing the generalized eigenvalue problem. This investigation enabled the identification of SWs modes of the system, tracked their variation in response to deformation and provided insight into their physical characteristics.

In a different study conducted by Farmer et al., they investigated the SWs dynamics of an antidot square lattice under Fibonacci distortion \cite{farmer2015magnetic}. This distortion was achieved by continuously applying the Fibonacci sequence along both orthogonal primitive vectors of a periodic square lattice as shown in the inset of figure \ref{fig:fib_dis}(d). The magnetization reversal of this structure shows the plateaus and step anomalies which may be considered due to the flux closure states. The calculated FMR spectra showed symmetry and reproducibility as shown in the figure \ref{fig:fib_dis}(d). Furthermore, it has been observed that despite the disruption of the fourfold symmetry in a finite periodic square antidot lattice caused by the Fibonacci distortion, the FMR data maintain a fourfold rotational symmetry concerning the direction of the applied DC magnetic field.

\section{Future Perspective}
\begin{figure}[!ht]
    \centering
    \includegraphics[width=0.8\linewidth]{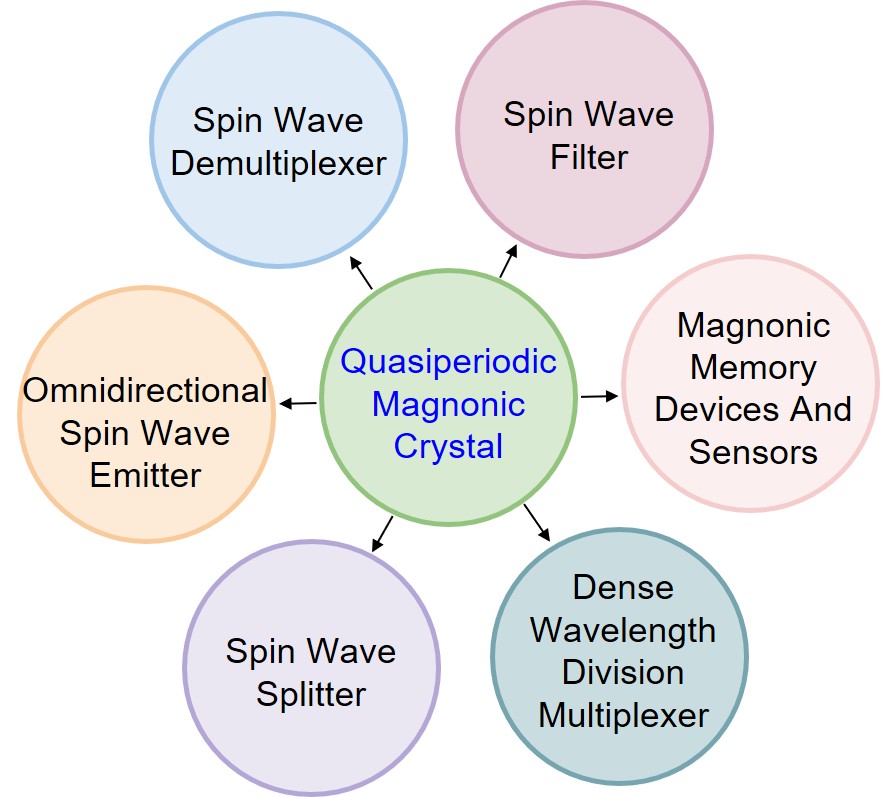}
    \caption{Schematic depicting the multifaceted potential applications of quasiperiodic MCs.}
    \label{fig:future_pros}
\end{figure}

In this article we have highlighted the key research directions that has been done so far in the field of quasiperiodic MCs. It has been found that quasiperiodic MCs, because of their unique feature of formation of mini frequency bands and localization of SWs modes, offer promising prospects for magnonic applications. The cartoon in figure \ref{fig:future_pros} depicts various proposed applications of quasiperiodic MCs in future magnonic devices. As demonstrated by Mehta et al. that the triangular fractal quasiperiodic MCs can efficiently serve as SWs filter and splitter through the opening of frequency gap with the fractal iteration \cite{mehta2023tunability}. A SW filter selectively permits the transmission of SWs within a specific frequency band while effectively blocking other frequencies. The SW splitter, on the other hand, splits a SW  mode to a multiple one. Lisiecki et al. in their work on Fibonacci quasiperiodic MC unveiled the formation of mini frequency gap and localization of SWs mode which again render their potential applications in SW filter and SW demultiplexer \cite{lisiecki2019magnons}. A SW demultiplexer functions by selectively sorting magnonic signals into distinct frequency components, providing efficient frequency-dependent separation. Another work on the quasiperiodic MCs in the form of Penrose tiling by Watanabe et al. depicts their potential application as an omnidirectional SWs emitter \cite{watanabe2021direct}. This tiling has the capability to emit SWs in any in plane direction driven by the unconventional symmetry of Penrose tiling. Furthermore, reference \cite{watanabe2020direct} shows the excitation of SWs on nanoholes based quasiperiodic MC characterized by tenfold rotation symmetry, such as the Penrose tiling. This excitation results in the formation of multiplexed magnonic nanochannels, illustrating their potential application as dense wavelength division multiplexer. Additionally, the work shown in reference \cite{park2021recursive} on ferromagnetic antidot fractal lattice with periodic boundary condition of square Bravais lattice shows the standing SW modes with fine localizations, offering precise and compact excitation, present themselves as a favourable candidate for a range of magnonic devices. These applications encompass memory devices and sensors, where pinpointed and space-efficient excitation is of paramount importance.

In order to incorporate magnonic nanodevices into magnonic networks, it's crucial to establish straightforward interconnections between the functional blocks situated within different layers of the entire magnonic circuit \cite{gubbiotti2019three,barman20212021}. The framework of 3D magnonics holds significant potential in this regard, presenting a multitude of possibilities for creating vertical connections. Furthermore, the advancements in nanofabrication methods hold the potential to create new quasiperiodic MC geometries in 3D, opening the door to the exploration of unprecedented emergent phenomena. SW dynamics and damping parameter \cite{vivas2012investigation} in 3D MCs in the form of meander-type waveguide \cite{beginin2019collective,beginin20183d,sadovnikov2022reconfigurable} and the sphere in hexagonal lattice \cite{krawczyk2010materials} has already been explored for their possible potential application in reconfigurable magnonic devices. Indeed, till now no work has been reported on the 3D quasiperiodic MCs \cite{socolar1986quasicrystals}. For periodic MCs, one approach to fabricating 3D structures is to manufacture layered system utilizing the same lithography methods used to create 2D systems \cite{lavrijsen2013magnetic}. Another approach to cultivating the 3D structures is through focused electron beam-induced deposition \cite{fowlkes2018high} or two-photon laser lithography \cite{williams2018two}.  These versatile fabricating strategies pave the way for the development of quasiperiodic MCs as well.  Exploring the SWs in 3D quasiperiodic MCs is important for their potential application in advanced magnonic devices.

The magnonic bandgap can be tuned in a periodic composite nanostructure made of two different materials, named as the bicomponent MCs. The SWs dynamics in bicomponent periodic MCs have been widely explored \cite{gubbiotti2012collective,choudhury2016shape}. Magnetization dynamics in bicomponent quasiperiodic MCs for some systems like Fibonacci sequence and Penrose tiling has also been studied \cite{rychly2015spin,rychly2018spin}. The similar idea can further be extended to explore the SWs dynamics in other bicomponent quasiperiodic MCs as well, considering their huge potential for the applications in future magnonic devices.

Dzyaloshinskii-Moriya interaction (DMI) is an asymmetric exchange interaction occurs at the interfaces of ferromagnets and heavy metals possessing strong spin-orbit interaction due to the lack of spatial inversion symmetry \cite{gallardo2019spin}. The DMI gives rise to various spin textures including spiral magnetic states, skyrmion lattices, and isolated skyrmions \cite{saha2019formation,feilhauer2020controlled}. Moreover, DMI imposes nonreciprocity in SW dispersion.
Investigating the impact of interfacial DMI on SWs dynamics of both single component and bicomponent MCs revels unidirectional propagation of the SWs \cite{ma2014interfacial}. The FMR spectra of 1D MCs and isolated stripes, influenced by DMI, exhibit the folding of magnonic bands in the first brillouin zones and the quantization of SWs \cite{mruczkiewicz2016influence}. Exploring the effect of  DMI on SWs dynamics of quasiperiodic MCs offers a pathway for potential advancements in future magnonic devices.

Recently, various physical phenomena occurring at the interfaces of magnetic heterostructures are being explored considering their potential for the development of energy efficient spintronic and magnonic devices. The voltage-controlled magnetic anisotropy (VCMA) \cite{rana2019towards} holds significant promise for advancing the field of low-power magnonic devices that can be entirely operated through electric field \cite{barman20212021,chumak2022advances,rana2017effect,miura2017voltage,rana2017excitation}. Wang et al. proposed a novel configuration involving a periodic arrangement of stripe-like 1D metallic gate electrodes on the top of MgO/Co structure to impose a periodic modulation of the perpendicular magnetic anisotropy (PMA) in Cobalt (Co) \cite{wang2017voltage}. This voltage-induced periodic modulation of PMA affects SWs propagation, leading to the formation of band gaps in the SWs spectrum. Notably, selectively applying PMA to specific gate stripes, rather than all, allows the selection of magnonic zone boundary. The formation of periodic magnonic nanochannels by VCMA has been demonstrated by Choudhury et al. \cite{choudhury2020voltage}. The SWs propagating through these nanochannels interact among themselves to form rich magnonic spectra, where band gap is tunable by the gate voltage. These concepts could be expanded by arranging the gate stripes in a quasiperiodic order, thereby forming voltage-controlled reconfigurable quasiperiodic MCs.

For the improvement of device functionality, it is also necessary to couple magnons with other quasiparticles such as photons, phonons.  The magnon-exciton coupling in various systems such as van der Waals heterointerface \cite{gloppe2022magnon}, antiferromagnetic semiconductor CrSBr has already been reported \cite{dirnberger2023magneto,bae2022exciton,diederich2023tunable}. Likewise, magnon-phonon coupling has been studied in magnetic insulators \cite{man2017direct}. Similar studies could be explored in quasiperiodic MCs composed of antiferromagnetic semiconductors and magnetic insulators for their effective utilization in hybrid devices.

The field of quasiperiodic MCs holds a multitude of exciting future prospects, firmly grounded in theoretical foundations. As technology continues to advance, it will provide inspiration to the field enabling the fabrication of diverse quasiperiodic magnonic geometries, including intricate 3D structures, their manipulation using various stimuli, especially electric field, strain and the exploration of potential magnonic devices.

\section{Summary}
In this review, we have provided an overview of the quasiperiodic MCs, which, in contrast to periodic MCs do not possess strict periodicity, and show quite complex and localized SWs spectra with a large number of band gaps. The magnetization dynamic of an artefact honeycomb and octagonal lattices show a rich SWs spectra compared to the periodic Bravais lattices. Moreover, the SWs spectra in the Fibonacci sequenced structure reveal a remarkably rich band pass structure with the self-similar behaviour of the magnonic bulk band in relation to the generalized Fibonacci generation number. Furthermore, the calculated IDOS of the quasiperiodic MCs in the form of Penrose and Ammann tiling shows the discernible plateaus within the spectra. These plateaus serve as indicators of frequency gaps, highlighting the presence of long-range order within the quasiperiodic MCs. The identification of localized SWs mode further underscores the quasiperiodicity of these magnonic tilings. Moreover, the amplitude distribution of magnetostatic SWs modes, exemplified in the Sierpinski square, follows a distinct geometrical scaling. The spatial profile of the SWs mode of a fractal structure at any iteration can be analysed using the mode profiles of the structure at the preceding iteration. The magnetization dynamics in Sierpinski triangle show the appearance of frequency gap with the iteration number exceeding a certain value and its tunability with the orientation and strength of the external magnetic field. In addition, the introduction of Fibonacci distortion in the Kagome artificial spin ice and square antidot lattice leads to the splitting of SWs modes into multiple branches, on top of that the SWs spectra can be tuned by altering the extent of lattice distortion. 

The quasiperiodic MCs landscape not only unveils the elegance of complex SWs spectra but also promises a technological renaissance. With potential applications covering filters, splitters, multiplexers, demultiplexers and more, these crystals will pave the way for innovative magnonic devices, heralding a transformative chapter in magnonics.

\section{Acknowledgement}
S S and R M  gratefully acknowledge the financial support of SERB with file Number SRG/2022/000191 and the Axis Bank Grant at
Ashoka University for the funding. R M acknowledges Ashoka University for senior research fellowship. B R acknowledges the financial support from NCN SONATA-16 project with Grant Number 2020/39/D/ST3/02378.

\newpage


\end{document}